\documentstyle[preprint,tighten,aps,psfig]{revtex}
\preprint{BNL-HET-98/30, TTP98-33, hep-ph/9809341}

\begin{document}

\title{Positronium hyperfine
splitting: analytical value at ${\cal O}(m \alpha ^6)$}

\author{Andrzej Czarnecki\thanks{
e-mail:  czar@bnl.gov}}
\address{Physics Department, Brookhaven National Laboratory,\\
Upton, NY 11973}
\author{Kirill Melnikov\thanks{
e-mail:  melnikov@particle.physik.uni-karlsruhe.de}}
\address{Institut f\"{u}r Theoretische Teilchenphysik,\\
Universit\"{a}t Karlsruhe,
D--76128 Karlsruhe, Germany}
\author{ Alexander Yelkhovsky\thanks{
e-mail:  yelkhovsky@inp.nsk.su }
}
\address{ Budker Institute for Nuclear Physics,         
\\
Novosibirsk, 630090, Russia}
\maketitle

\begin{abstract}
We present an analytic calculation of the
${\cal O}(m\alpha^6)$ recoil corrections to the
hyperfine splitting (HFS) of the ground state
energy levels in positronium. We find
$\Delta E_{\rm rec} = m\alpha^6
               \left( - \frac{ 1 }{ 6 } \ln \alpha + \frac{331}{432}
               - \frac{ \ln 2 }{ 4 }
               - \frac{17\zeta(3)}{8\pi^2} +
               \frac{5}{12\pi^2} \right) \approx
m\alpha^6 \left( - \frac{ 1 }{ 6 }
\ln \alpha +  0.37632 \right)$, confirming Pachucki's numerical result
\cite{Ph}.
We present a complete analytic formula for the
${\cal O}(m\alpha^6)$ HFS of the positronium ground state and,
including ${\cal O}(m\alpha^7 \ln^2\alpha)$ effects, find
$
\left ( E(1^3S_1)-E(1^1S_0) \right )_{\rm theory}
= 203\; 392(1)\; {\rm MHz}$.
This differs from the experimental results by about 3 standard
deviations.
\end{abstract}


Spectroscopy of positronium, an atom consisting of an electron and a
positron, provides a sensitive test of the Quantum Electrodynamics
(QED) applied to bound state problems.  Electron and positron are
so much lighter than the lightest hadrons that the effects of strong
interactions are negligible compared with the accuracy of present and
any conceivable future experiments.  For this reason positronium
represents a unique system which can, in principle, be described with
very high precision by means of the QED only. One should also mention
that the measurements of positronium spectrum are performed with
very high accuracy \cite{Mills88}.

There are two
main approaches used in the studies of bound states.  The
Bethe--Salpeter method is based on an exact two-body relativistic wave
equation \cite{BR}. The other approach is the so--called
Non-Relativistic Quantum Electrodynamics (NRQED) \cite{CL}, which is
an effective field theory based on the QED for small energies and
momenta. Thus, by construction, the NRQED takes advantage of
non--relativistic energy of the electron and positron in positronium.

We note in passing that similar techniques are used nowadays
for describing  heavy quark--antiquark bound states.
 From this perspective, positronium may serve as a testing ground for
methods which can in the future be applied to the QCD.

The HFS of the positronium ground state (i.e. the difference between
the energies of the ground state with total spin $1$ and $0$)
belongs to one of the most accurately measured physical quantities.
Two experimental values of the highest precision are:
\begin{equation}
\Delta \nu \equiv  E(1^3S_1)-E(1^1S_0)
=203\;387.5(1.6)\;{\rm MHz},
\label{Mills}
\end{equation}
and
\begin{equation}
\Delta \nu =203\;389.10(0.74)\;{\rm MHz},
\label{Hughes}
\end{equation}
obtained respectively in \cite{Mills,Ritter}.  The bulk of this effect
is of the order $m\alpha^4$, where $m=0.51099907(15)$ MeV
(corresponding to $1.2355898(4)\times 10^{14}$ MHz) is the
electron mass and $\alpha =1/137.03599959(51)$ is the fine structure
constant.  Higher order corrections must be included to fully exploit
the experimental accuracy.  In particular,  since
$m\alpha^6 = 18.658\; {\rm MHz}$, a complete calculation
at this order is required.
With an exception of the leading logarithm, effects of  
order $m\alpha^7$ have not yet been studied.  Clearly, the experimental
precision warrants further studies of such corrections.

The history of theoretical calculations of various
contributions to the HFS of positronium is quite long.  They can be
represented by a series in powers and logarithms of the fine structure
constant,
\begin{equation}
\Delta \nu = m\alpha^4 \left( n_0+ \alpha n_1 + \alpha^2 n_2
+\ldots\right).
\end{equation}
The leading order ${\cal O}(m\alpha^4)$ HFS was obtained in
\cite{Pir,Ber,Fer},   
\begin{equation}
n_0  = {7\over 12}.
\end{equation}
The first correction was calculated
in \cite{KK},
\begin{equation}
n_1  = -{1\over \pi}\left( {8\over 9} + {\ln 2\over 2}\right).
\end{equation}
The second correction consists of the following contributions:
\begin{equation}
m\alpha^6 n_2  = \Delta E_{g-2}+\Delta E_{\rm annih}
+\Delta E_{\rm rad\;rec}+\Delta E_{\rm rec}.
\end{equation}
The logarithmic contributions at this order, 
${\cal O}(m\alpha^6 \ln\alpha)$, present in the annihilation $\Delta
E_{\rm annih}$ and recoil corrections $\Delta E_{\rm rec}$, were found
first \cite{BY,CLlog}.  $\Delta E_{g-2}$ arises from the anomalous
magnetic moment of the electron at ${\cal O}(\alpha,\alpha^2)$.  The
three-, two- and one-photon annihilation contributions giving $\Delta
E_{\rm annih}$ were found in \cite{ABZ,AAB,HL}, respectively. The
non-annihilation radiative recoil contributions $\Delta E_{\rm rad\;
rec}$ were studied in \cite{STY,PhK}, while pure recoil corrections
$\Delta E_{\rm rec}$ were discussed in \cite{Ph,CL,AS}.

\begin{figure} 
\hspace*{-2mm}
\begin{minipage}{16.cm}
\vspace*{3mm}
\[
\hspace*{-5mm}
\mbox{ 
\begin{tabular}{cc}
\psfig{figure=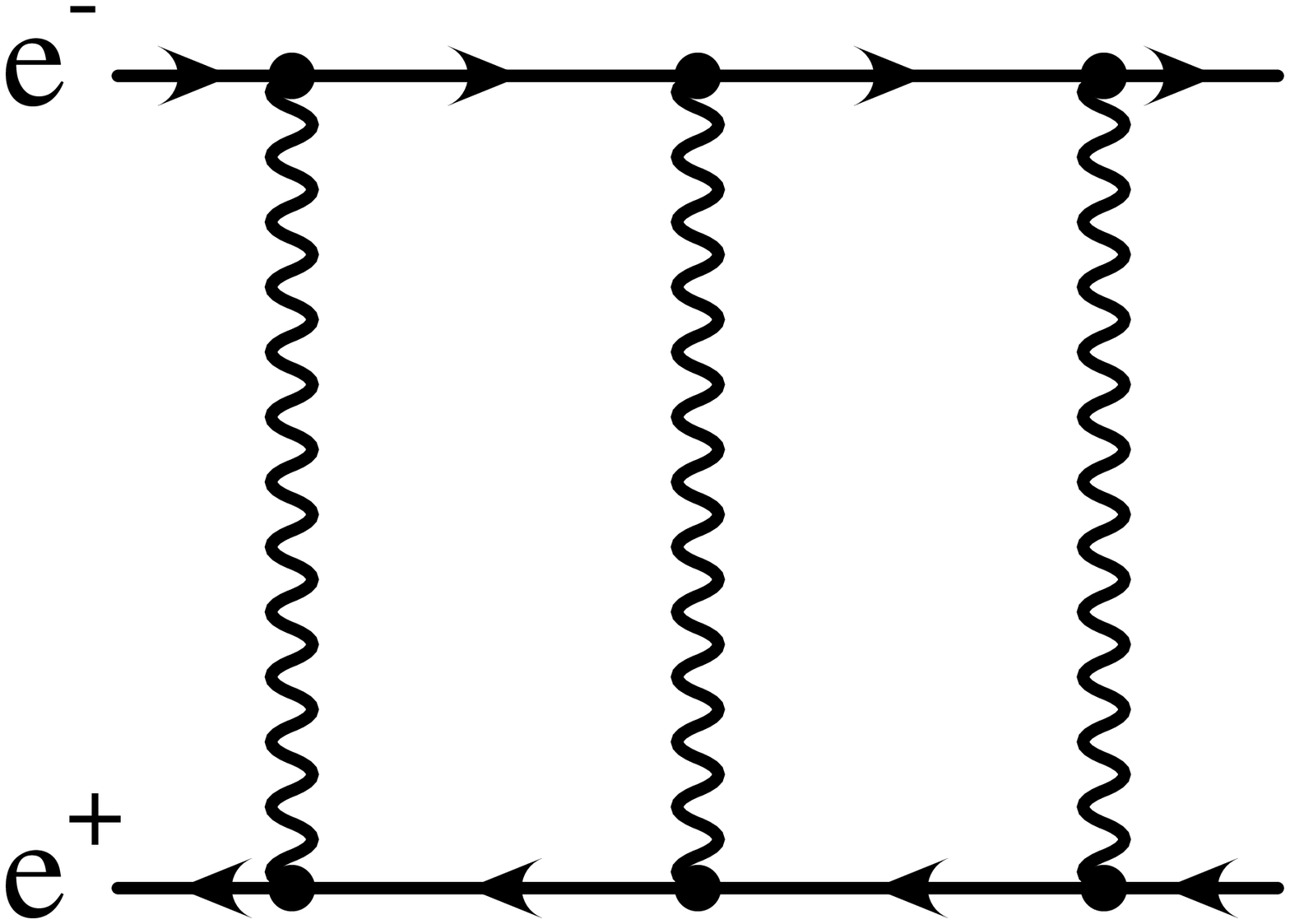,width=30mm,bbllx=72pt,bblly=291pt,%
bburx=544pt,bbury=530pt} 
& \hspace*{10mm}
\psfig{figure=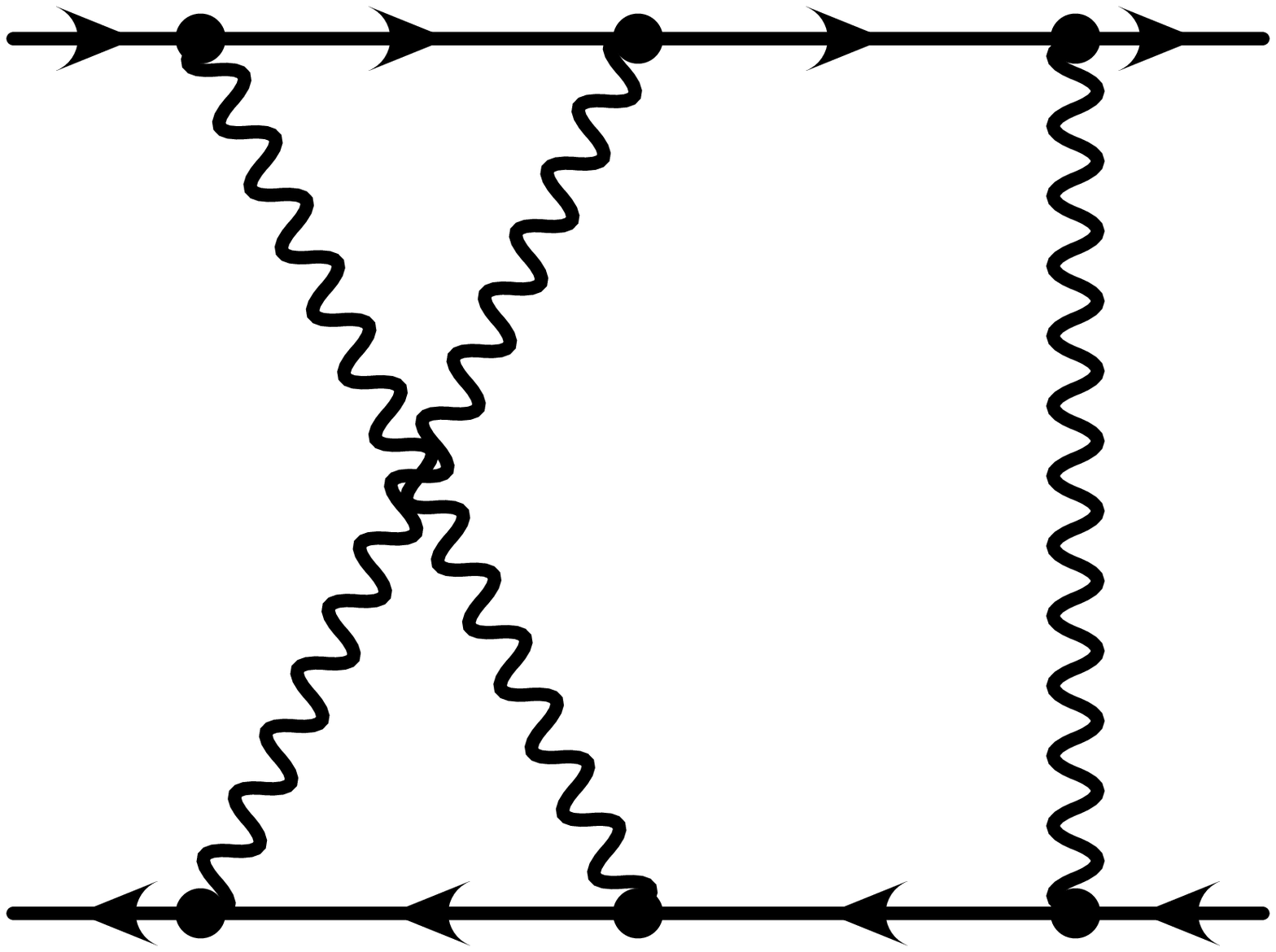,width=30mm,bbllx=72pt,bblly=291pt,%
bburx=544pt,bbury=530pt} 
\\[12mm]
\psfig{figure=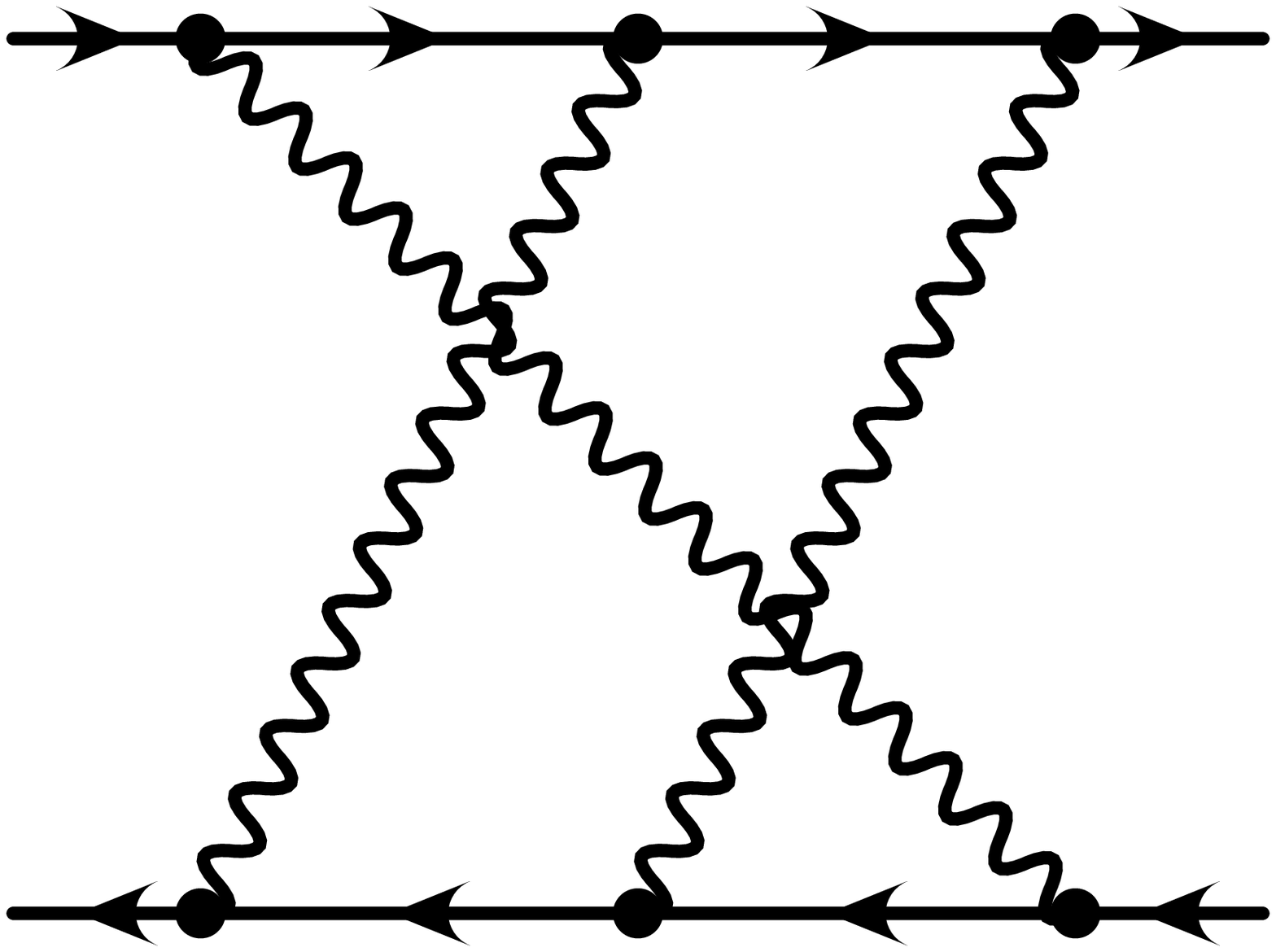,width=30mm,bbllx=72pt,bblly=291pt,%
bburx=544pt,bbury=530pt} 
& \hspace*{10mm}
\psfig{figure=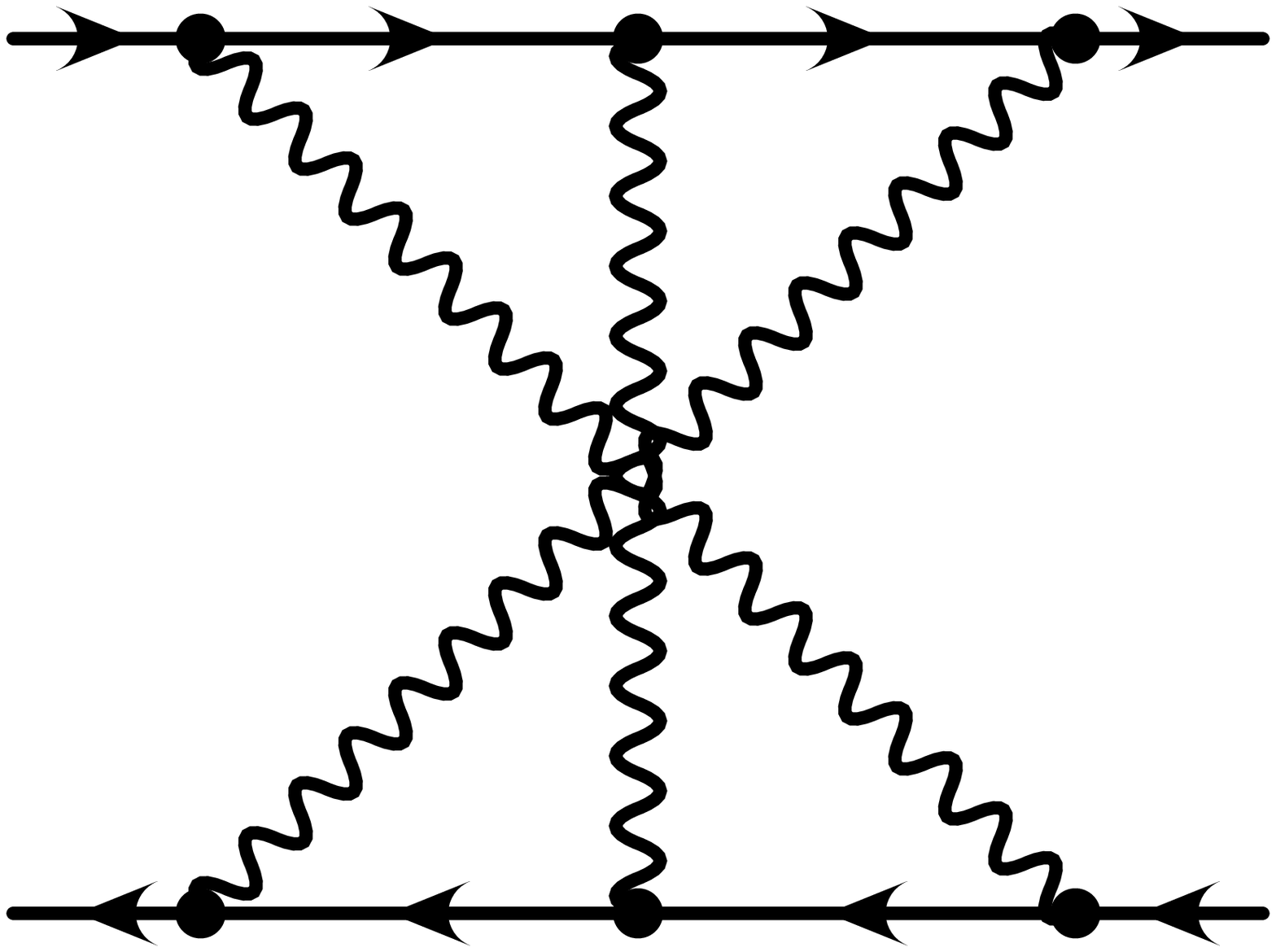,width=30mm,bbllx=72pt,bblly=291pt,%
bburx=544pt,bbury=530pt} 
\end{tabular}
}
\]\vspace*{6mm}
\end{minipage}
\caption{Feynman diagrams representing pure recoil corrections to
positronium HFS.}
\label{fig1}
\end{figure}

For most of these contributions, several independent calculations were
performed and an agreement was achieved.  Moreover, the results for
all contributions to HFS are known in the analytic form, with the
exception of the pure recoil corrections $\Delta E_{\rm rec}$.  
By pure recoil  
corrections one understands those induced by diagrams where each
virtual photon is created by electron and absorbed by positron, as
shown in Fig.~\ref{fig1}.   For
these effects,  three independent calculations arrived at
three different results \cite{Ph,CL,AS}. 
The discrepancy has not been
clarified so far and the resulting uncertainty in the theoretical
prediction for the HFS of the ground state is much
larger than the experimental error. The importance of clarifying this
theoretical point has been emphasized by several authors.  In this
Letter we present an analytic calculation of these corrections.
Numerically our result coincides with
Ref.~\cite{Ph}.  

We start with a short description of the framework of our calculation,
leaving the details to a separate publication.
First, we calculate the on--shell scattering amplitude for
non--relativistic $(v \ll 1)$ particles to the necessary order. Along
with the leading amplitude of a  single Coulomb exchange, it includes
the relative order ${\cal O}(v^2)$ Breit corrections and also higher
order ${\cal O}(v^4,\alpha v^3)$ terms. By construction, it is
gauge--invariant.  Taken with a minus sign, this amplitude provides
the effective potential for non--relativistic particles.

Further, we solve the Schr\"odinger equation incorporating
corrections to the Coulomb potential using ordinary quantum mechanical
perturbation theory.  According to standard rules, we get the ${\cal
O}(m\alpha^6)$ correction to the ground--state energy as a sum of the
relativistic corrections to the tree level and one--loop scattering
amplitude, and of the second order correction due to the Breit potential.
Previously, this scheme was used for the calculation of the
${\cal O}(m\alpha^6 \ln\alpha)$ corrections to the levels of $S$-states
\cite{KMYlog} and of the ${\cal O}(m\alpha^6)$ corrections to the levels
of $P$-states \cite{KMYp}.

An implementation of this program leads to a divergent result.  The
reason for this divergence is well known -- it is the application of
the non--relativistic expansion in the relativistic momentum region,
where it is not appropriate.  The divergence is canceled
if one includes additional short-distance or hard--scale
contributions to the scattering amplitude, which cannot be obtained 
from the non--relativistic expansion.

To deal with the divergences in both the non--relativistic region and
in the short--distance corrections we employ dimensional
regularization. In the context of bound state calculations in QED this
regularization scheme was used in Ref.~\cite{PiSo}, where the known
results for ${\cal O}(m\alpha^5)$ corrections to positronium energy
levels were successfully reproduced.  The advantage of the dimensional
regularization is that it makes the matching calculation of the
low-scale effective theory and the complete QED extremely
simple. To obtain the contribution of a given
Feynman diagram to the Wilson coefficient of the $\delta (\vec{r})$-like
effective operator, we only need to calculate that diagram for zero
incoming momenta of all particles.  We stress that this is only
correct if one uses dimensional regularization for both infrared and
ultraviolet divergences.  With any other regularization scheme an
additional calculation is required.

We find that in the sum of the short- and long--distance
contributions the singularities $1/\epsilon=2/(4-D)$ disappear
and one arrives at a finite result.

Since the dimensional regularization is used throughout the Letter,
we mention how the spinor algebra was treated.
To calculate the shift in the ground state energy
due to some operator ${\cal O}_i$ one has to calculate
the trace of the form
$\mbox{Tr} \left [ \Psi^\dagger {\cal O}_i \Psi \right ]$,
where $\Psi$ is an appropriate wave function.
The spinor parts of the relevant wave functions are:
$$
\Psi_{\mbox p} = \frac{1+\gamma_0}{2\sqrt{2}} \gamma_5,\qquad
\Psi_{\mbox o} = \frac{1+\gamma_0}{2\sqrt{2}} \mbox{\boldmath {$ \gamma $}}
\mbox{\boldmath {$\xi $}}  ,
$$
for para- and orthopositronium states, respectively.  In the latter
case
$\mbox {\boldmath {$ \xi$}}$ is the 
polarization vector. The traces are calculated in the $D$-dimensional
space.  Since we always encounter an even number of $\gamma_5$'s, we
treat them as anticommuting. We also average over directions of the
vector $\mbox{\boldmath {$\xi$}}$.  In order to obtain 
corrections to the HFS we
first calculate separately the traces for ortho- and parapositronium
states and then take the difference of the two.

The problem is naturally divided up into the calculation of the matrix
elements of the effective operators (soft contributions) and the
Wilson coefficients of the effective $\delta (\vec{r})$-like 
operators (hard
contributions) in the effective Hamiltonian:
\begin{equation}
\Delta E_{\rm rec} =\Delta_{\rm soft} E_{\rm rec} +
\Delta_{\rm hard} E_{\rm rec} .
\end{equation}
 The calculation of
Wilson coefficients is always done for the incoming and outgoing
particles at rest. Both technically and conceptually, this
is close to the calculation of the matching coefficient of the
vector quark-antiquark current in QCD and its NRQCD counterpart,
described e.g.~in \cite{BSS,CzM}.

\begin{figure} 
\hspace*{-2mm}
\begin{minipage}{16.cm}
\vspace*{3mm}
\[
\hspace*{-5mm}
\mbox{ 
\begin{tabular}{cc}
\psfig{figure=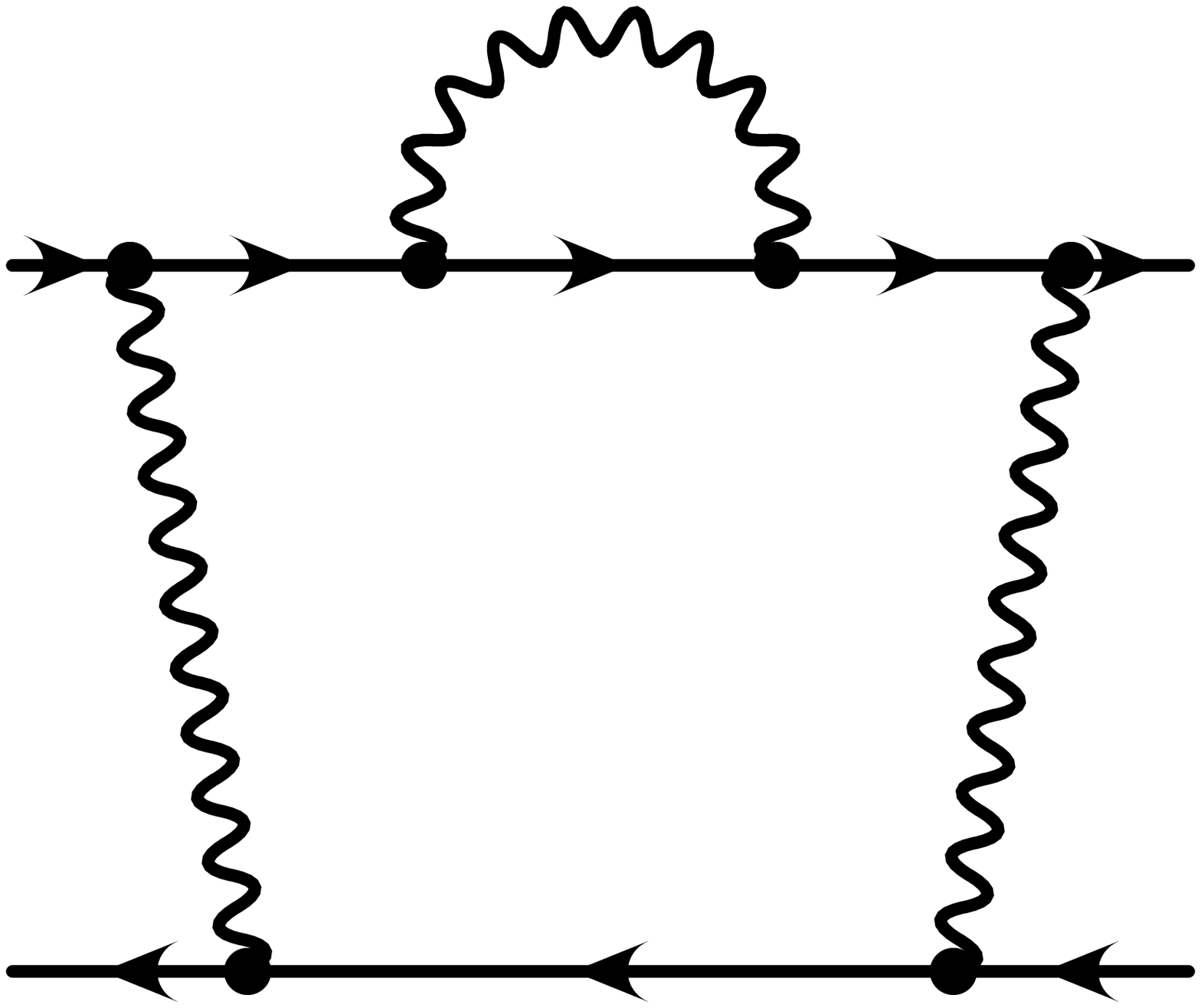,width=30mm,bbllx=72pt,bblly=291pt,%
bburx=544pt,bbury=530pt} 
& \hspace*{10mm}
\psfig{figure=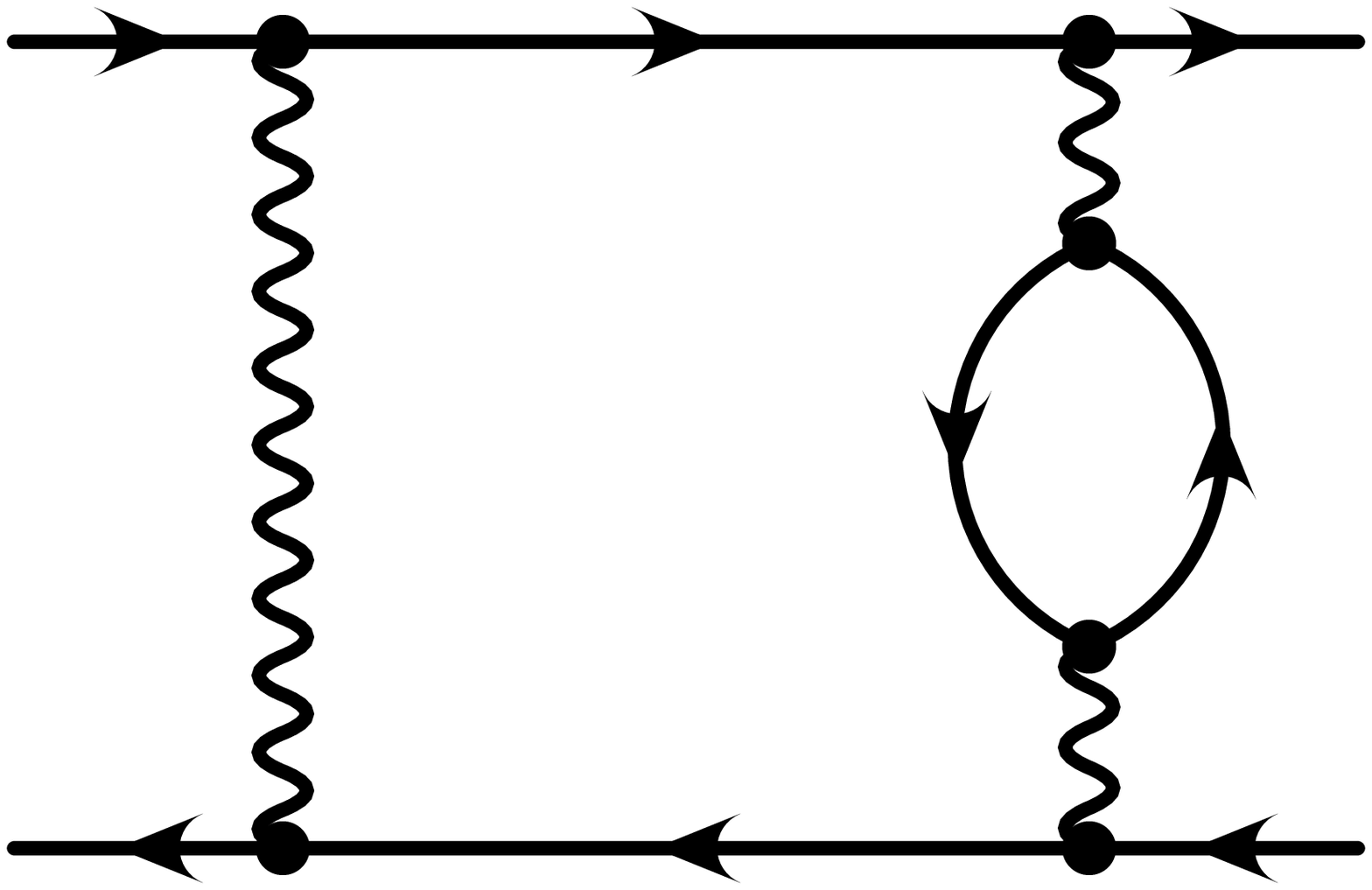,width=30mm,bbllx=72pt,bblly=291pt,%
bburx=544pt,bbury=530pt} 
\\[12mm]
\psfig{figure=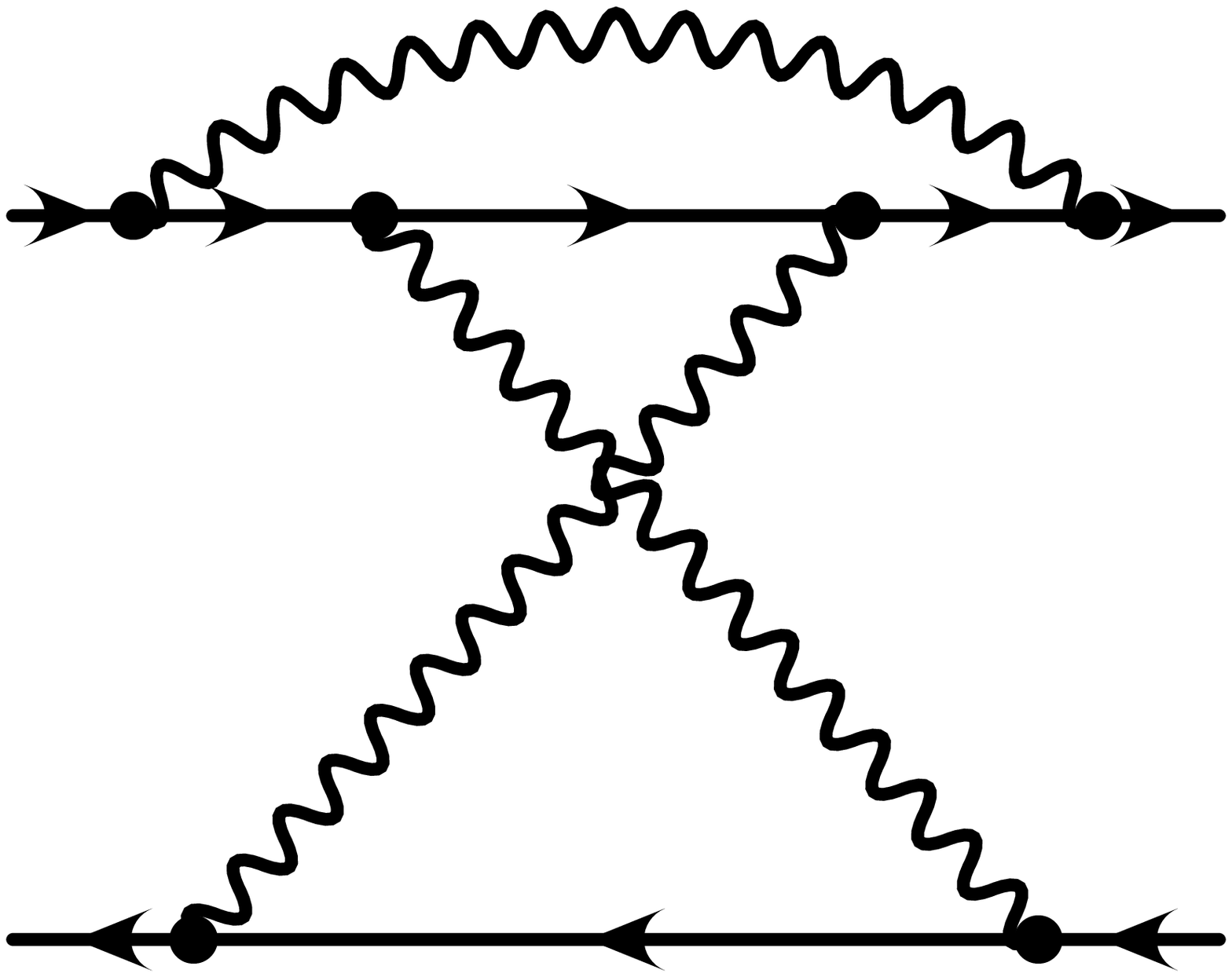,width=30mm,bbllx=72pt,bblly=291pt,%
bburx=544pt,bbury=530pt} 
& \hspace*{10mm}
\psfig{figure=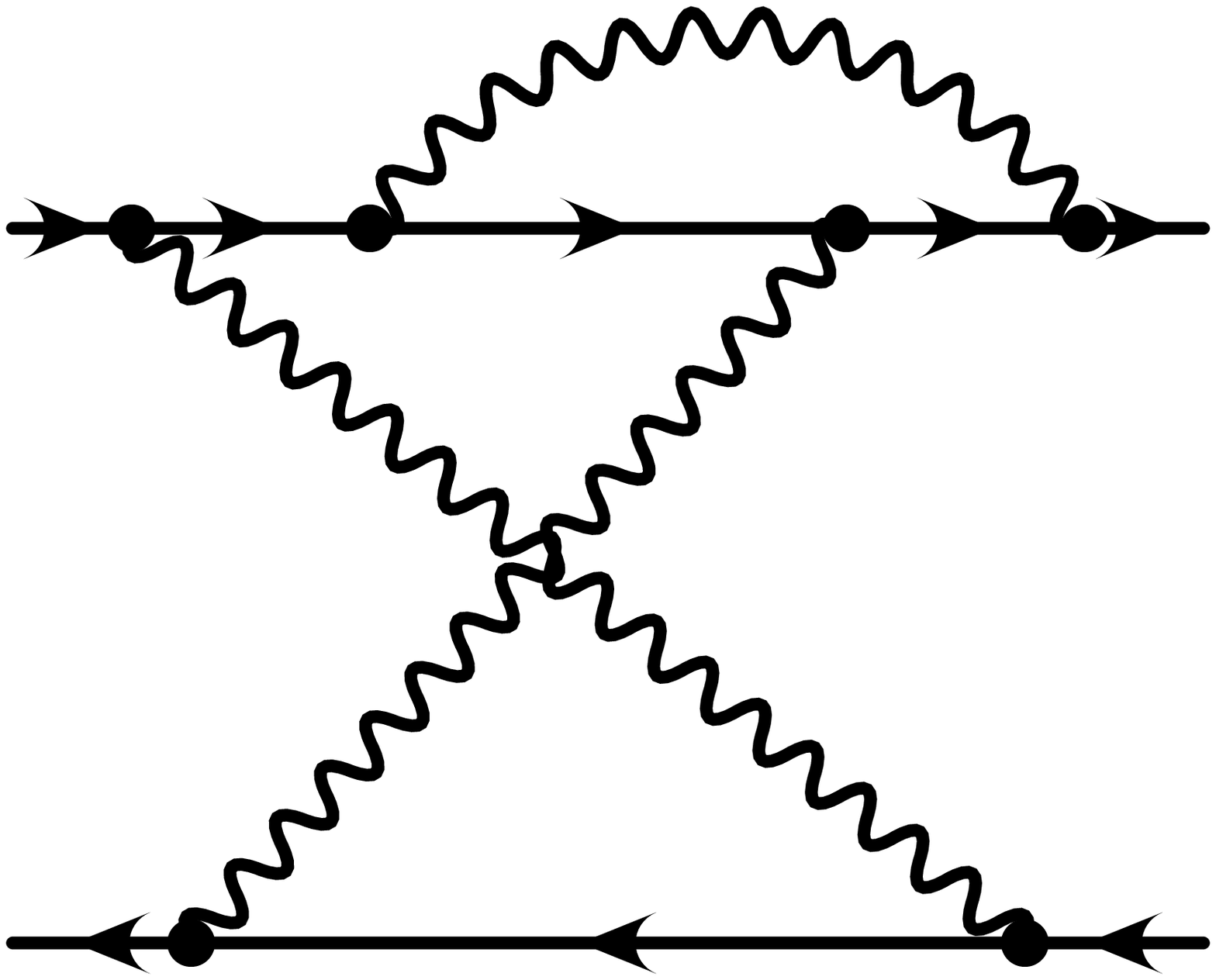,width=30mm,bbllx=72pt,bblly=291pt,%
bburx=544pt,bbury=530pt} 
\end{tabular}
}
\]\vspace*{6mm}
\end{minipage}
\caption{Examples of radiative recoil corrections to
positronium HFS.}
\label{fig2}
\end{figure}

This technique is
remarkably useful for the so--called radiative
recoil corrections to the HFS, where
one of the  three exchanged photons is created and absorbed
by the same particle, as shown in Fig.~\ref{fig2}. 
It is
sufficient to calculate the corresponding integrals exactly at the
threshold in dimensional regularization, since there are no
non--relativistic contributions to the radiative recoil corrections
and no matching is required. Performing this calculation we obtain:
\begin{equation}
\Delta E_{\rm rad\; rec}
= m\alpha^6 \left( \frac{\zeta(3)}{2\pi^2} +\frac{4}{3}\ln 2
-\frac{79}{48}+\frac{41}{36\pi^2} \right).
\end{equation}
This result is in complete agreement with the analytic result
published previously \cite{PhK}.

Applying the same technique to
obtain the hard--scale contribution to  the recoil corrections
$\Delta E_{\rm rec}$ we obtain:\footnote{We neglect factors
$\Gamma^2(1+\epsilon)$ 
and $(4\pi)^{2\epsilon}$ which do not contribute
to the final, finite result.}
\begin{equation}
\Delta_{\rm hard} E_{\rm rec}
 = \frac{ \pi\alpha^3 }{ 3 m^2 } |\psi_d(0)|^2
               \left( - \frac{ 1 }{\epsilon } + 4 \ln m
               - \frac{51\zeta(3)}{\pi^2} +
               \frac{10}{\pi^2} - 6\ln 2 \right).
\label{hard}
\end{equation}
In the above equation $\psi_d(0)$ stands for the value at the origin
of the ground state solution of the $d$--dimensional Schr\"odinger
equation.

The calculation of the soft scale contributions requires
the treatment of the relativistic corrections to the tree level
and one-loop scattering amplitudes, as well as the second iteration
of the Breit potential. The main difficulty associated with this
calculation is that it should be done in $d=3-2\epsilon$ dimensions,
thus necessarily spoiling some simplifying features of the Coulomb
problem in three dimensions. Still, the calculation is feasible.
Since the non--relativistic Hamiltonian is only singular
for $r \to 0$, it turns out possible to extract this divergence
in the form $|\psi_d(0)|^2/\epsilon$,  without solving the
Schr\"odinger equation in $d$-dimensions.
Our final result for all non--relativistic contributions reads
\begin{equation}
\Delta_{\rm soft} E_{\rm rec}
 = \frac{ \pi\alpha^3 }{ 3 m^2 } |\psi_d(0)|^2
               \left( \frac{ 1 }{\epsilon } - 4 \ln(m\alpha) +
               \frac{331}{18} \right).
\label{soft}
\end{equation}

In the sum of the hard and non-relativistic
contributions, Eqs.~(\ref{hard},\ref{soft}), the $1/\epsilon$ divergences
disappear and we can take the limit $\epsilon \to 0$ in the sum.
We thus arrive at the final result for the  recoil
corrections to the HFS of the positronium ground state:
\begin{equation}
\Delta E_{\rm rec}
 = \Delta_{\rm hard} E_{\rm rec} + \Delta_{\rm soft} E_{\rm rec}
 = m\alpha^6
               \left( - \frac{ 1 }{ 6 } \ln \alpha + \frac{331}{432}
               - \frac{ \ln 2 }{ 4 }
               - \frac{17\zeta(3)}{8\pi^2} +
             \frac{5}{12\pi^2} \right).
\label{recoil}  
\end{equation}
Numerically this is $\Delta E_{\rm rec} = m\alpha^6 \left( - \frac{ 1
}{ 6 } \ln \alpha + 0.37632 \right)$ which is in excellent agreement
with Ref.~\cite{Ph}, where for the non-logarithmic part of the 
correction a number 0.3767(17) was obtained.
In view of the fact that in Ref.~\cite{Ph} a
different regularization was used, this agreement gives us confidence
in the correctness of the result\footnote{When this work was
completed, we were informed \cite{AB} about an independent numerical
calculation of the recoil corrections. Though that work is still
in progress, its preliminary results seem to coincide with the
results of Ref.~\cite{Ph} and of the present work with rather
good accuracy.}.

The recoil correction was the last correction
to positronium bound state HFS  not known analytically. 
Having obtained its value (Eq.~(\ref{recoil})), we are now in position
to present the final analytic result for the HFS of the positronium
ground state including ${\cal O}(m\alpha^6)$ terms:
\begin{eqnarray}
\lefteqn{E(1^3S_1)-E(1^1S_0)
= m\alpha^4 \left \{ \frac {7}{12} -
\frac{\alpha}{\pi} \left (\frac {8}{9}+\frac{1}{2} \ln 2 \right )
\right.} \nonumber \\
&& \qquad \qquad \left.
+\frac {\alpha^2}{\pi^2} \left [
- \frac{5}{24}\pi^2 \ln \alpha
+ \frac {1367}{648}-\frac{5197}{3456}\pi^2
 + \left ( \frac{221}{144}\pi^2 +\frac {1}{2} \right ) \ln 2
-\frac{53}{32}\zeta (3) \right ] \right \}.
\label{hfsfin}
\end{eqnarray}
Numerically this corresponds to 
$\Delta \nu = 203\; 392.899\; {\rm MHz}$, if we use the central values
for $m$ and $\alpha$, given after Eq.(\ref{Hughes}).

To arrive at the final prediction for the HFS splitting of the
positronium ground state, one should try to quantify the theoretical
error. The errors caused by the uncertainties in the fine structure
constant and the electron mass are $\sim 0.003$ and $0.07$ MHz, respectively. 
The main uncertainty comes from the unknown higher order effects. 
Though formally
$m\alpha^7 \sim 0.1\; {\rm MHz}$, the leading  ${\cal
O}(m\alpha^7 \ln^2 \alpha)$ terms contribute $-0.92\; {\rm MHz}$ to the HFS
\cite{DL}.  Therefore, it remains very important to calculate the
remaining, non-leading terms in ${\cal O}(m\alpha^7)$. In this context
we note that the complete ${\cal O}(m\alpha^6)$ correction, including
the $m\alpha^6\ln \alpha$ term, gives a shift of $11.79\; {\rm MHz}$,
whereas the term $m\alpha^6\ln \alpha$ alone contributes $19.12\; {\rm
MHz}$.  We see that an estimate of the complete correction based on
the $m \alpha^6 \ln(\alpha)$ approximation would not be accurate.

At the moment the best we can do is to take the leading log
contribution ${\cal O}(m\alpha^7 \ln^2 \alpha)$ as an estimate of the 
higher orders corrections to
HFS and thus attribute $\sim 1\; {\rm MHz}$ uncertainty to the
theoretical prediction.  Combining this theoretical uncertainty with
Eq.~(\ref{hfsfin}), we obtain the theoretical estimate for the
HFS of the ground state of the positronium:
\begin{equation}
\Delta \nu _{\rm theory}
= 203\; 392(1)\; {\rm MHz}.
\end{equation}
Compared to the experimental results Eq.~(\ref{Mills},\ref{Hughes})
we observe a significant deviation of the order of $3-4$ experimental
errors.  We look forward to future improved measurements of
positronium HFS and their confrontation with QED.

We are grateful to A. Burichenko for informing us about his
results prior to publication.
K.M. would like to thank the High Energy Theory Group at the
Brookhaven National Laboratory for hospitality
during the final stage of this project.
This research was supported in part by  DOE under grant
number DE-AC02-98CH10886, by BMBF under grant number BMBF-057KA92P,
by Gra\-duier\-ten\-kolleg ``Teil\-chen\-phy\-sik'' at the
University of Karlsruhe and by the  Russian Foundation for Basic Research
under grant number 98-02-17913.


\end{document}